# Network traffic prediction based on ARFIMA model

Dingding Zhou[1],Songling Chen[1] ,Shi Dong[2,3]

[1] Department of Laboratory and Equipment Management, Zhoukou Normal University, Zhoukou, 466001, China

[2] School of Computer Science and Technology, Zhoukou Normal University, Zhoukou, 466001, China

[3] School of Computer Science and Engineering, Southeast University, Nanjing, 211189, China

## Abstract

ARFIMA is a time series forecasting model, which is an improve d ARMA model, the ARFIMA model proposed in this article is d emonstrated and deduced in detail. combined with network traffi c of CERNET backbone and the ARFIMA model,the result sho ws that,compare to the ARMA model, the prediction efficiency a nd accuracy has increased significantly, and not susceptible to sa mpling.

## 1. Introduction

Network traffic prediction is an important application of network management and network measurement research direction, network traffic prediction model is the model to predict trends of sometime things in the future [1-3] under the theory guidance. Network traffic model roughly is divided into two categories, traditional traffic model (the Poisson model, the Markov model, the waterfall model, AR model, ARMA [4-6] model). The second category is new traffic model, such as wavelet analysis, neural networks, fuzzy theory, and chaos theory and support vector machine. General auto regressive (AR) or auto regressive moving average (ARMA) is applied in traffic prediction; paper [11] introduced some method which can be applied to traffic prediction. Dong yan etc.al [7] proposed the network traffic prediction based on the flourier model, and prediction function of the model was given. Literature [8-9] proposed prediction model of regression summation moving average (ARIMA) model which is the improvement of the ARMA model using the differential method to make the model more stationary sequences than traditional prediction model of network The flow forecast better reflect the network characteristics of the self-similar figures, unexpected continuity. However, the model can only predict the short-term network traffic. Prediction precision also has relatively larger improvement space. This paper deeply study on ARFIMA [10] model and adopts ARFIMA model to predict real trace records and netflow sampling flow records. The results show that the ARFIMA can get more effective network traffic prediction.

## 2. ARFIMA Model Overview

If $\{y_t\}$ is a stationary process and satisfies the differential equation $\varphi(L)(1-L)^{\eta}(1-L)^{d}y_t=\theta(L)\mu_t$ , then the $\{ y_t \}$ is ARFIMA (p, $\eta$ , d, q) process. Where L is the causal operator, let $\mu_t$ as a white noise sequence, $\eta$ is the integer-order differential operator, d is fractional difference operator, and -0.5<d<0.5, and $\varphi(L)$ and $\theta(L)$ were polynomial causal operator of stable p and q order and root is outside of the unit circle. Obviously, a necessary and sufficient condition to the process $\{ y_t \}$ for the ARFIMA (p, $\eta$ , d, q) process is $(1-L)^{\eta}(1-L)^{d}y_t$ for ARMA (p, q) process. Steps of ARFIMA (p, $\eta$ , d, q) model building: The first step: analysis of the sequences in long-term memory factors and determine the value of d;
Second step: Fractional differential and obtains zero mean ARMA (p, q) sequence;
Third step: ARFIMA (p, $\eta$ , d, q) given order, i.e. determine p and q values.

### 2.1 R/S analysis process

Hurst exponent is a parameter proposed by the water expert H.E.Hurst in the mid 20th century, which can judge whether discriminant time series exist time dependent. He found this period often lasts for years by observing river flooding and dry season. This is different from the past, in theory, that the annual water flow is independent and identically distributed Gaussian variables, is also different from the traditional assumptions of the markov chain.On the basis of experience, this paper adopts a new hurst parameter to analyze the time series cluster phenomenon, and proposes the R/S method which can calculate the hurst parameters.

l) The return series of length N $\{ip_t = p_{t+1} - p_t\}$ is defined, and it is divided into the A consecutive subintervals of length n. Each subinterval is marked as $Ia, a = 1,...,A$. Thus in each point can be expressed as $Ip_{k,a}$ , $k = 1,...,n, a = 1,...,A.$

2) For each sub-interval $Ia$ of length n, calculate the mean as: $e_a = \frac{1}{n}\sum_{k=1}^{n} ip_{k,a}$

3) For single subinterval, the cumulative mean deviation is calculated as $y_{k,a}$ . $y_{k,a} = \sum_{i=1}^{k}(ip_{i,a} - e_a), k = 1,...n$ . From the above equation we can see, cumulative sum of single subinterval mean difference sequence $\{y_{1,a}, y_{2,a}, ...y_{n,a}\}$ is zero.

4) Range of single subinterval is defined as

$R_{la}$ =Max( $y_{k,a}$ )−Min( $y_{k,a}$ ), $k = 1,...n$ 。

5) The standard deviation of each sub-interval is calculated

$$Sl_a = \sqrt{\frac{1}{n}\sum_{k=1}^{n}(E_{k,a} - e_a)}^2$$

And use it to rescale/standardize range $(R_{la}/S_{la})$ , according to the changes scope divided by the standard deviation of the observed values and obtain a dimensionless quantity, which will make different sequences with comparability. So through partition of length n, we can calculate average scaling range of A sub-the interval:

$$(R/S)_n = \frac{1}{A}\sum_{a=1}^{A}\frac{R_{la}}{S_{la}}$$

6) According to hurst proposed model, and there is a relational expression $(R/S)_n = (c*n)^H$ which is taken logarithm on both sides, thus $\log(R/S)_n = H*\log n + \log c$ . $R_{\ln}$ is a range of deviation mean accumulated value on n data in a time sequence, called range of n data, represents the maximum variation range of time series; $S_n$ shows the standard deviation of the time series; represents the degree of deviation from the mean, is metric of divergence degree. (R/S) represents the size of range can be evaluated by $S_n$ , which is applicable to a random process of long-term memory function. H is the hurst exponent, c is a constant. Using least-squares regression method, H values and their standard deviation can be obtained. According to the research of Ellis C, calculated the fractional difference d of ARFIMA model and it can be solved by the Hurst exponent, that is d = H-0.5.

## 2.2 Fraction gradient of derivation process

For the original sequence $\{y_t\}$ can be formulated to be a d-order differential, and where L is a causal operator. Differential sequence denoted $\{W_t\}$ . In order to facilitate the calculation, let $y_0 = 0, W_t = (1-L)^d y_t$ , in order to fractional differential, first consider the binomial expansion of $(1-L)^d$ , and obtain the following formula(1)

$$(1-L)^d = 1 - dL + \frac{d(d-1)L^2}{2} - \frac{d(d-1)(d-2)L^3}{3!} + ... \quad (1)$$

In formula (1), for any real number with D>-1 can be represented as used a hyper geometric function.

$$(1-L)^d = \sum_{k=0}^{\infty}\frac{\Gamma(d+1)}{\Gamma(k+1)\Gamma(d-k+1)}(-1)^k L^k \quad (2)$$

When the d value is determined, the function $\sum_{k=0}^{\infty}\frac{\Gamma(d+1)}{\Gamma(k+1)\Gamma(d-k+1)}(-1)^k$ is just a function of k, it denoted g(k) .The formula can be expressed as $W_t = (\sum_{k=0}^{\infty} g(k)L^k)y_t$ .

Formula also can be expressed as:

When t = 0, $W_0$ = 0.

When t = 1, $W_1 = g(0)y_1 + g(1)y_0 = g(0)y_1$ = 0.

When t = 2, $W_2 = g(0)y_2 + g(1)y_1$ .

… …

When t = N, $W_N = g(0)y_N + g(1)y_{N-1} + ... + g(N)y_1$ .

The above calculation process can be expressed by matrix $W = [W_1.W_2....W_N], y = [y_1.y_2...y_N]$

$$G = \begin{bmatrix} g(0)g(1)g(2)...g(N-1) \\ .......g(0)g(1)g(2)..g(N-2) \\ ...............g(0)g(1) \\ ....................................g(2) \\ ....................................g(1) \\ ....................................g(0) \end{bmatrix}$$

The fractional differential will be completed by calculating $W = y*G$ .

## 2.3 ARFIMA $(p,\eta,d,q)$ model predictive formula derivation

ARFIMA $(p,\eta,d,q)$ model

$\varphi(L)(1-L)^\eta(1-L)^d y_t = \theta(L)\mu_t$ can also be represented

$(1-\varphi_1 L - \varphi_2 L_2 - ... - \varphi_p L^p)(1-L)^d x_t$

$= (1-\theta_1 L - \theta_2 L^2 - ... - \theta_q L^q)\mu_t$

Where $X_t = (1-L)^d y_t$ then

$$X_t = \varphi_1 x_{t-1} + \varphi_2 x_{t-2} + ... + \varphi_p x_{t-p} + \frac{(1-\theta_1 L - \theta_2 L2 - ... - \theta_q L^q)\mu_t}{(1-L)^d}$$

According to formula (2), $(1-L)^d = \sum_{k=0}^{\infty} \frac{\Gamma(d+1)}{\Gamma(k+1)\Gamma(d-k+1)}(-1)^k L^k$ is referred to as $(\sum_{k=0}^{\infty} g(k))L^k$ , the above formula can be expressed as

$$x_t = \varphi_1 x_{t-1} + \varphi_2 x_{t-2} + \ldots + \varphi_p x_{t-p} + \frac{\mu_t^2}{(\sum_{k=0}^{\infty} g(k))L^k \mu_t} - \ldots - \frac{\theta_q \mu_{t-q}^2}{(\sum_{k=0}^{\infty} g(k))L^k \mu_{t-p}}$$

The denominator is obtained by the fractional differential for $\mu = (\mu_1 \mu_2 \cdots \mu_{t-q})$ sequence, so the sequence will be referred to as $fd = (fd(t)\, fd(t-1) \ldots fd(t-q))$ , then above formula is simplified as

$$x_t = \varphi_1 x_{t-1} + \varphi_2 x_{t-2} + \ldots + \varphi_p x_{t-p} + \frac{\mu_t^2}{fd(t)} - \ldots - \frac{\theta_q \mu_{t-q}^2}{fd(t-q)}$$

Let $g_{T+h}$ as minimum mean square error prediction, and it satisfies the following formula

$$E(x_{T+j} \mid x_T, x_{T-1}, \ldots) = \begin{cases} x_{T+j}, & j \le 0 \\ g_{T+j}, & j > 0 \end{cases}$$

$$E(\mu_{T+j} \mid \mu_T, \mu_{T-1}, \ldots) = \begin{cases} \mu_{T+j}, & j \le 0 \\ 0, & j > 0 \end{cases}$$

According to above formula, we can draw a prediction value of minimum mean square error.

$$x_t = \phi_1 x_{T+h-1} + \phi_2 x_{T+h-2} + \ldots + \phi_p x_{T+h-p} + \frac{\mu_{T+h-1}^2}{fd(T+h)} - \ldots - \frac{\theta_q \mu_{T+h-q}^2}{fd(t-q)}$$

## 3. Experiments and results analysis

### 3.1 Experimental data

Data is collected from a node CERNET backbone in a three-day October 2009, consecutive 72-hour sfrom 0:00 to 24:00 .Detail traffic data are as follows:

Table 1: First day traffic data

| | | | | | | | | |
|---|---|---|---|---|---|---|---|---|
| time | 0 | 1 | 2 | 3 | 4 | 5 | 6 | 7 |
| traffic | 4579980 | 3710512 | 3157889 | 2853392 | 2605959 | 2460145 | 2343611 | 2568329 |
| time | 8 | 9 | 10 | 11 | 12 | 13 | 14 | 15 |
| traffic | 3689533 | 5435866 | 6889225 | 7014512 | 7773121 | 8224229 | 8247586 | 8380588 |
| time | 16 | 17 | 18 | 19 | 20 | 21 | 22 | 23 |
| traffic | 8809613 | 8555500 | 8354096 | 9362043 | 9356253 | 9219356 | 8560516 | 6369390 |

Table 2: Second day traffic data

| | | | | | | | | |
|---|---|---|---|---|---|---|---|---|
| time | 0 | 1 | 2 | 3 | 4 | 5 | 6 | 7 |
| traffic | 4584550 | 3754345 | 3190912 | 2784617 | 2534228 | 2351439 | 2246381 | 2441217 |
| time | 8 | 9 | 10 | 11 | 12 | 13 | 14 | 15 |
| traffic | 3313881 | 4798367 | 6268091 | 6830835 | 7915579 | 8692846 | 8831177 | 8985398 |
| time | 16 | 17 | 18 | 19 | 20 | 21 | 22 | 23 |
| traffic | 9190998 | 8732356 | 8664592 | 9249369 | 9250066 | 9038823 | 8453056 | 6545901 |

Table 3: Third day traffic data

| | | | | | | | | |
|---|---|---|---|---|---|---|---|---|
| time | 0 | 1 | 2 | 3 | 4 | 5 | 6 | 7 |
| traffic | 4761961 | 3787515 | 3149364 | 2802171 | 2576379 | 2397031 | 2336571 | 2629421 |
| time | 8 | 9 | 10 | 11 | 12 | 13 | 14 | 15 |
| traffic | 3685423 | 5216799 | 6561625 | 7156008 | 7831581 | 8010846 | 7921752 | 8037197 |
| time | 16 | 17 | 18 | 19 | 20 | 21 | 22 | 23 |
| traffic | 8402647 | 8262956 | 8393854 | 9000149 | 9132337 | 9097003 | 8504556 | 6380823 |

The above data is analyzed by using the MATLAB tools, and finally hurst index value is 0.8745 which is obtained. Oberviously, due to 0.5<0.875<1, research on characteristics of self-correlation we can see, the traffic data meet characteristics of self-correlation.

## 3.2Network traffic prediction

In order to be able to predict the result of evaluation, need to introduce the following evaluation metric, root mean square error (RMSE)

$$RMSE = \sqrt{\frac{\sum_{X=1}^{n}(X_k - \overline{X_k})^2}{n}} \qquad (3)$$

Relative root mean square error （RRMSE)

$$RRMSE = \sqrt{\frac{\sum_{X=1}^{n}(\frac{X_k - \overline{X_k}}{X_k})^2}{n}} \qquad (4)$$

Theoretically, less the error is and more accurate prediction is, therefore the smaller two evaluation metrics and it will be expressed the better.

## 3.3 Simulation results

This paper adopts three models to forecast the traffic data provided in the 3.1 section, forecasting results based on the traditional time series model and ARFIMA model are shown in table 4:

Table 4: Forecasting results of three traffic models

| NO | Traffic Model | RMSE | RRMSE |
|---|---|---|---|
| 1 | ARMA(0,1) | 0.3157 | 0.4306 |
| 2 | ARIMA(0,1,1) | 0.3642 | 0.4385 |
| 3 | ARFIMA(1,1.3544,1) | 0.2235 | 0.3364 |

From the above experiment results of three traffic models we can see ARFIMA model adopting R/S parameter analysis method have made a very good improvement in both RMSE and RRMSE, prediction error is relatively small, and prediction accurate is relatively high.

When the day's traffic is divided by every 5 minutes time granularity, total one day has 288 data. It will be expressed as T5m.

When the day's traffic is divided by every 1 hour time granularity, total one day has 24 data. It will be expressed as T1h.

Similarly, the day's traffic is divided by every 2 hours as a time granularity division, a total of one day has 12 data. It will be expressed as T2h.

We use ARFIMA model to forecast traffic time sequence data of these three kinds of different granularity, the results are shown in the following table:

Table 5: Forecasting results of different time size

| Time granularity | Traffic Model | RMSE | RRMSE |
|---|---|---|---|
| T5m | ARFIMA(1,1.3544,1) | 0.3423 | 0.4508 |
| T1h | ARFIMA(1,1.3544,1) | 0.3789 | 0.4578 |
| T2h | ARFIMA(1,1.3544,1) | 0.2345 | 0.3476 |

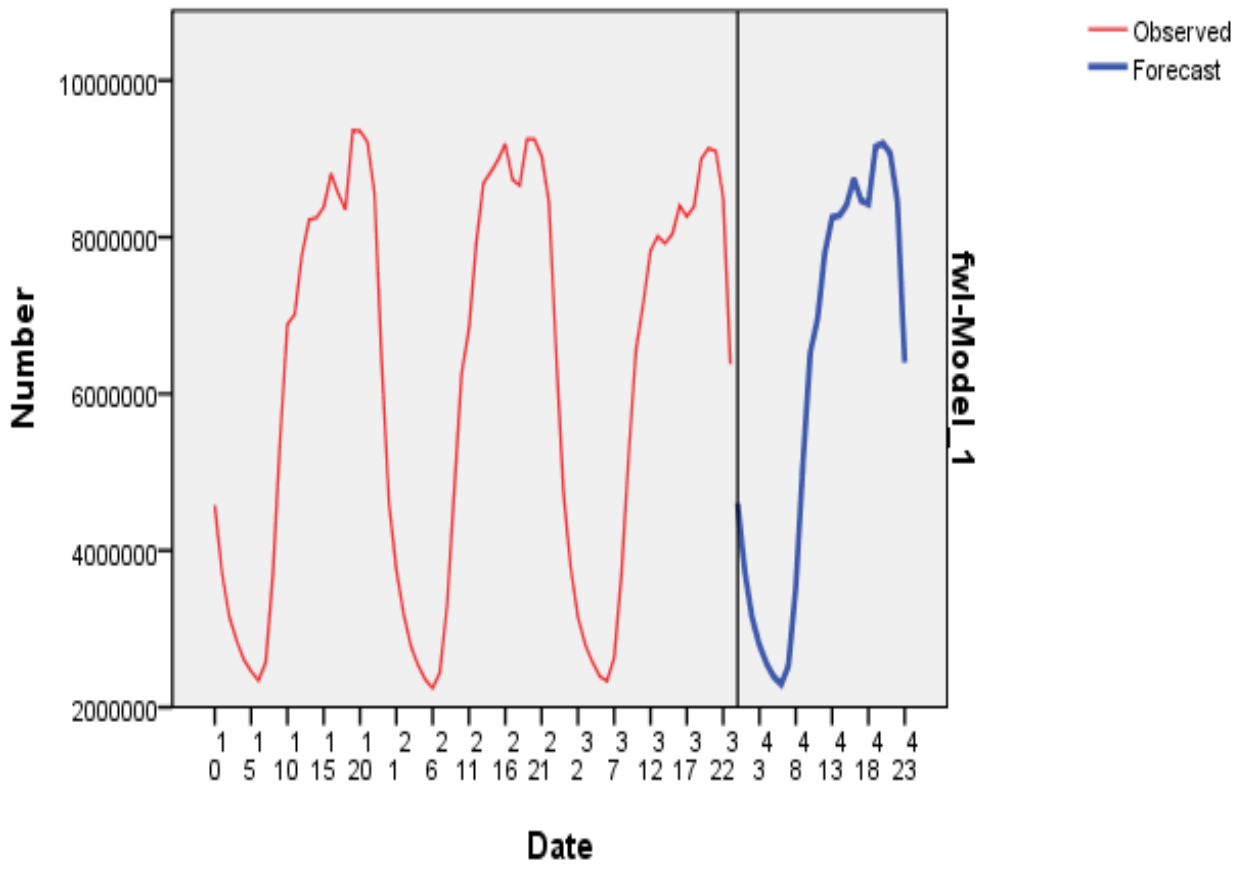

Fig. 1 Results of observed and forecast.

From table 5 we can see, the smaller the time size, the greater the prediction error. Forecasting precision and accuracy of 5 minutes time granularity is the lowest. Fig. 1 shows the comparison of 5 minutes size observed value and the predicted value. In order to further research, we have analyzed and compared other traditional time series models ARMA and ARIMA, the results are shown in Fig.2. We can see ARFIMA prediction is more close to the original traffic flow, the difference of prediction results is maximum in ARIMA model and the original traffic flow.

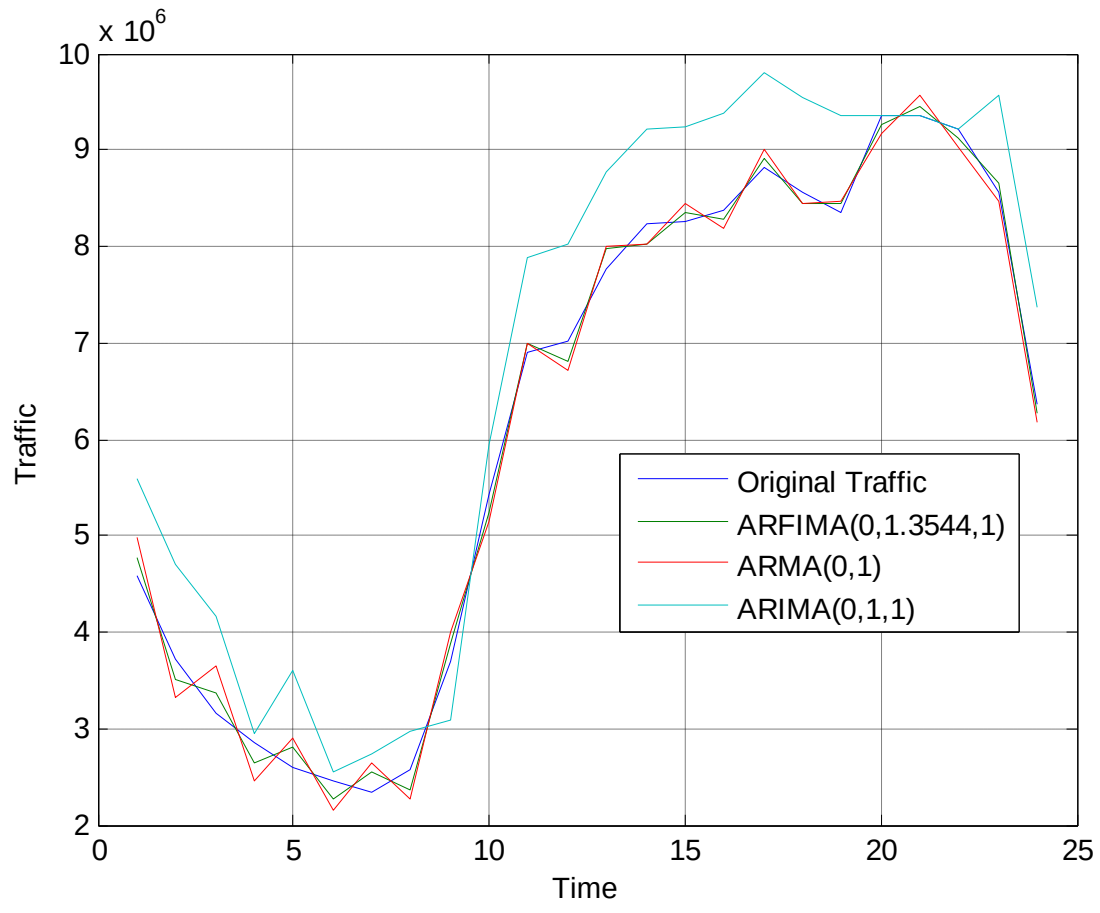

Fig. 2 Comparison of forecasting model and original traffic

The current sampling has been widely applied to network management, and impact of the sampling size on flow prediction results assessment has not been studied, so this paper adopts systematic sampling method for packet sampling. Systematic sampling proportion is in accordance with the 1024:1; 256:1; 64:1; 8:1; 1:1. After sampling the experiment results is shown as following:

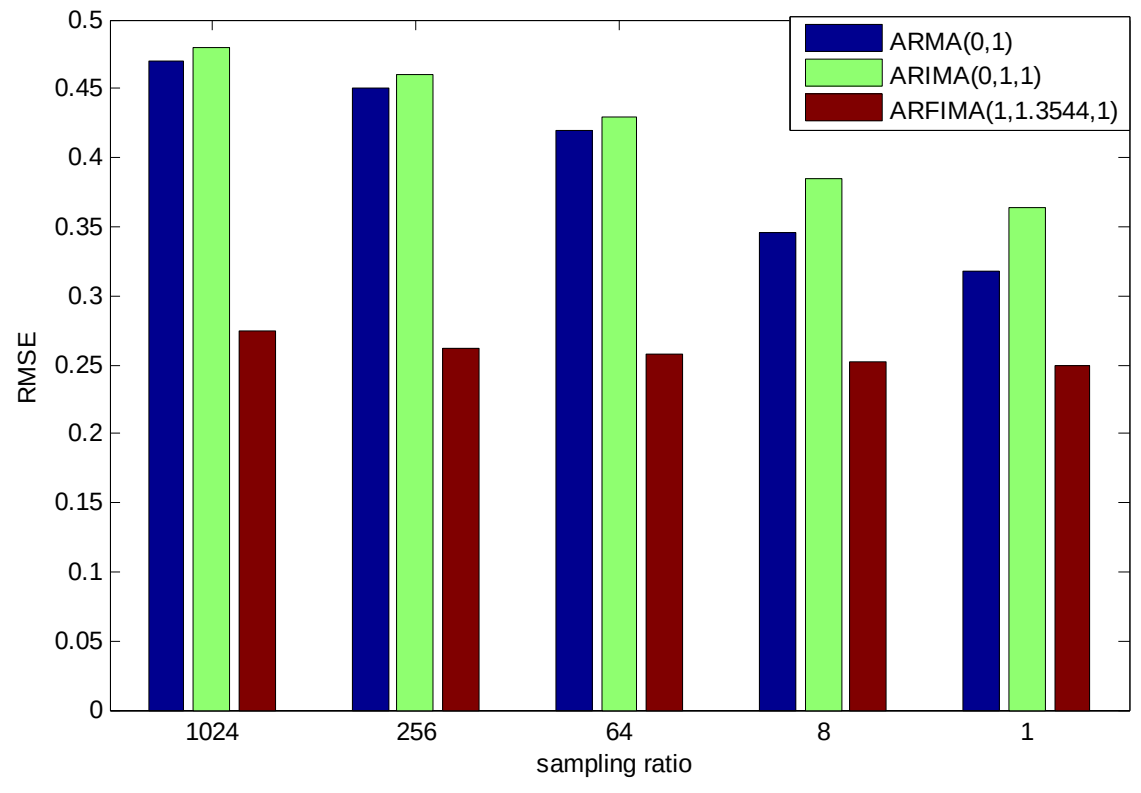

Fig. 3 RMSE in different sampling ratio

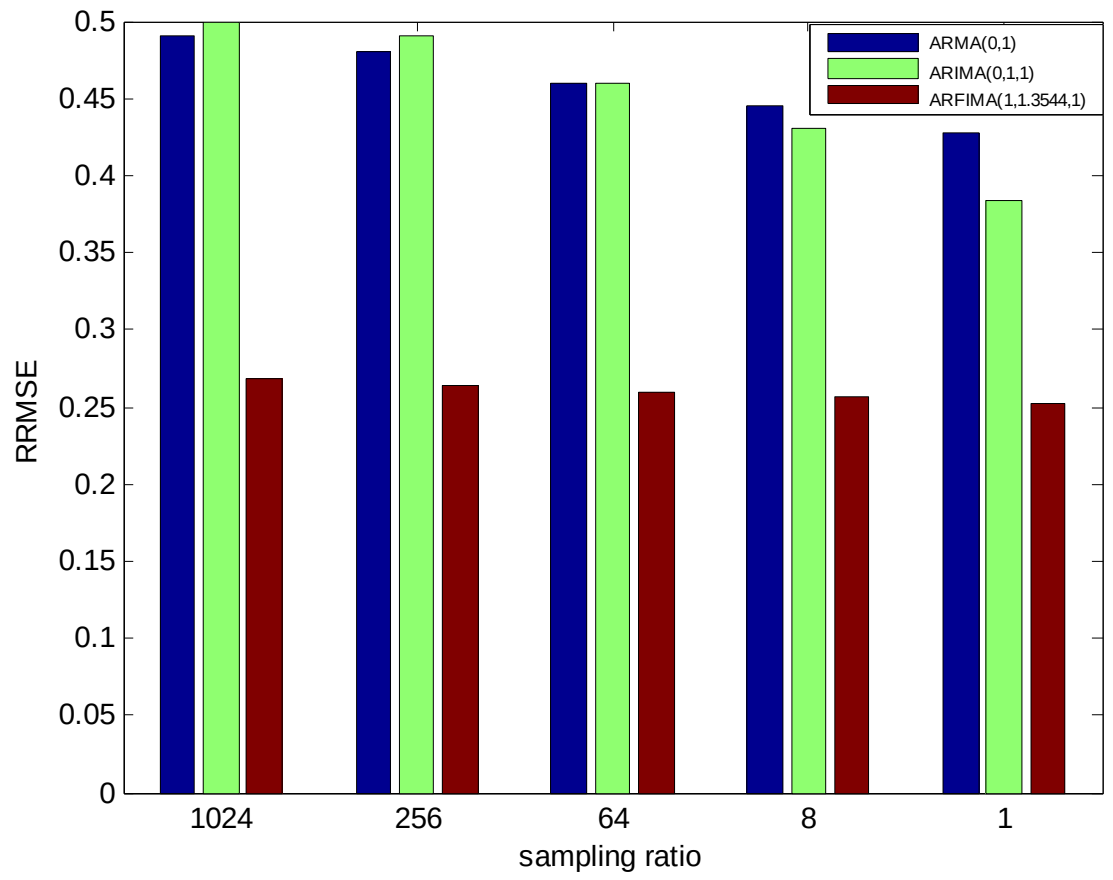

Fig.4 RRMSE in different sampling ratio

Fig 3 and 4 show error analysis of three models in different sampling ratio, with the sampling ratio decreasing, the prediction error is reduced in different degree, but compared with ARFIMA, ARMA and ARIMA decreased greatly, and sampling sensitivity of ARMA and ARIMA is larger, and the prediction error change of ARFIMA is not very obvious, the sampling sensitive of ARFIMA is less, it is more beneficial to research and analysis of long-range dependence. Based on this conclusion, we can apply ARFIMA model to network management service prediction module with sampling mechanism, which can more accurately predict changes of network traffic.

## 4. Conclusions

Network traffic forecasting in the network management and network operation get extensive application, through the analytical research of the present network traffic forecasting traditional model, this paper proposes ARFIMA model, and analyzes and derives ARFIMA model theory, and adopts ARFIMA model to study and analyze traffic data from China northeast network center, the results were predictable, based on three different time granularity division we use modeling and analysis method to study traffic data, experimental results show the ARFIMA model is better than the traditional time series models in the long correlation analysis. Considering the impact of sampling packets in current network management with sampling mechanism, analyzes the model under different sampling ratio, the results show that the ARFIMA model can more effectively carry out the network prediction, and less affected by sampling.

### Acknowledgments

This paper is supported by National 973 Plan Projects (2009CB320505) and National Science and Technology Plan Projects (2008BAH37B04).

## References

[1]Phillips, P. C. B., "Non-stationary Time-series Analysis and Cointegration", Oxford University Press, 1994

[2]Box, G. "P & GM Jenkins. Time Series Analysis: Forecasting and Control", Francisco:Holden-Day, 1976

[3]Litterman,R.B.(1986). "Forecasting with Bayesian vector autoregressions: Five years of experience" Journal of Business & Economic Statistics, Vol.4, No.1, 1986, pp.25-38

[4]Monahan, J. F. "Fully Bayesian analysis of ARMA time series models", Journal of Econometrics, Vol.21, No.3, 1983, pp. 307-331

[5] Xiong, Y. and D. Y. Yeung. "Mixtures of ARMA models for model-based time series clustering", in: Proceedings of the IEEE International Conference on DateMining, Maebashi city, Japan, 9-12 December2002，PP.717-720

[6]Kleibergen, F. and H. Hoek. "Bayesian analysis of ARMA models using noninformative priors", Erasmus School of Economics (ESE).1995

[7] Dong Sun Enchang, Sun Yanhua, "network traffic forecast based on Fourier model best", Application Research of Computers, Vol.4, No.1, 2010, pp.1419-1421.

[8]Kalpakis,K., D. Gada, et al. (2001). "Distance measures for effective clustering of ARIMA time-series", in: proceedings of the IEEE International Conference on DateMining，SanJose， CA.USA， 29November-2December 2001， pp.273-280.

[9]YU G Q,ZHANG C H. "Switching ARIMA model based forecasting for traffic flow", Proc of ICASSP. Piscataway: IEEE, 2004, pp.429-432.
[10]Ravishanker, N. and B. K. Ray, "Bayesian prediction for vector ARFIMA processes." International Journal of Forecasting, Vol.18, No.2, 2002, pp. 207-214
[11]Bazghandi, A , "Techniques, Advantages and Problems of Agent Based Modeling for Traffic Simulation", International Journal of Computer Science,Vol.9,No.1,2012,pp.115-119

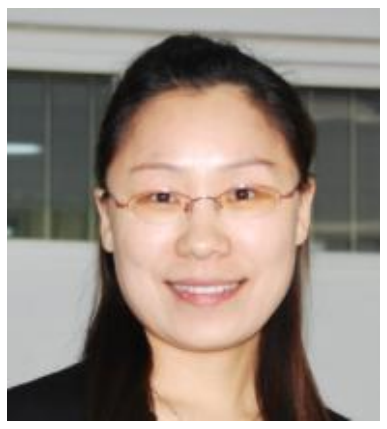

**Dingding Zhou** was born in Zhoukou, China, on February, 1980. She received the Computer Science B.S degree in Henan university. Nowadays she is lecturer in zhoukou normal university. Major research interests are in high speed communications, mobility, security and QoS guarantees.
.

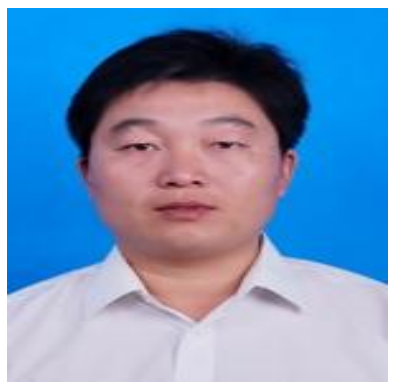

**Songling Chen** was born in Zhoukou, China, on February, 1974. he received the Computer Science B.S deg in Zhengzhou university. Nowadays he is lecturer in zhoukou normal university. His major research interests are in high speed communications, mobility, security and QoS guarantees.

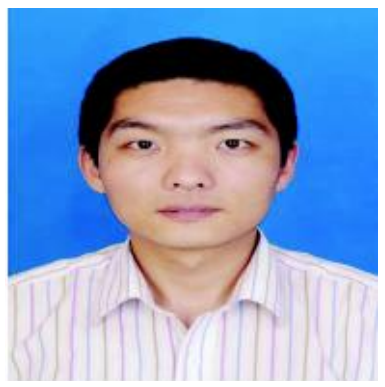

**Shi Dong** was born in Zhoukou, China, on NOV 05, 1980. He received the Master degree in Computer science from University of Electronic and technology of China. In 2009, nowadays he is pursuing for PhD degree in school of computer science from southeast university. Major interests are network management, network security